\documentclass[lettersize,journal]{IEEEtran}
\usepackage{amsmath,amsfonts}
\usepackage{algorithmic}
\usepackage{algorithm}
\usepackage{array}
\usepackage[caption=false,font=normalsize]{subfig}
\usepackage{textcomp}
\usepackage{stfloats}
\usepackage{url}
\usepackage{verbatim}
\usepackage{graphicx}
\usepackage{cite}

\usepackage{comment}
\usepackage{hyperref}
\usepackage[dvipsnames]{xcolor}

\newcommand*{\rev}[1]{{\color{black} #1}}

\begin{document}

\title{Rethinking Beam Management: Generalization Limits \\Under Hardware Heterogeneity}

\author{Nikita Zeulin, Olga Galinina, Ibrahim Kilinc, Sergey Andreev, and Robert W. Heath Jr.
\thanks{Nikita Zeulin, Olga Galinina, and Sergey Andreev are with Tampere University, Tampere, Finland (emails: name.lastname@tuni.fi).}%
\thanks{Ibrahim Kilinc and Robert W. Heath Jr. are with University of California San Diego, CA, USA (emails: \{ikilinc, rwheathjr\}@ucsd.edu).}
}

\maketitle

\begin{abstract}
Hardware heterogeneity across diverse user devices poses new challenges for beam-based communication in 5G and beyond. This heterogeneity limits the applicability of machine learning (ML)-based algorithms.
This article highlights the critical need to treat hardware heterogeneity as a first-class design concern in ML-aided beam management. We analyze key failure modes in the presence of heterogeneity and present case studies demonstrating their performance impact. Finally, we discuss potential strategies to improve generalization in beam management. 
\end{abstract}


\section{Introduction}
The diversity of user devices in 5G cellular networks and beyond creates a fundamental challenge for multiple-input multiple-output (MIMO) communication. Beam management becomes harder to scale in high-frequency bands such as the 7--24 GHz range (FR3) expected in 6G. To accelerate beam management procedures, ML-based approaches can leverage past data and side information, such as position, orientation, LIDAR measurements, or images.
A major limitation of most prior ML-related work is the assumption of homogeneity in deployment scenarios. Analogous to the classic ML risk of training and testing on similar data distributions, in beam management, training and testing on scenarios with only a single type of base station (BS), user, and environment may lead to overfitting and poor generalization to real-world deployments~\cite{akrout2023domain}. 

Beam-based communication systems are profoundly heterogeneous. 
By heterogeneity, we here refer to the presence of multiple diverse configurations, discrete as the number of antenna elements or continuous as the antenna orientation or location.  
This heterogeneity arises from multiple sources, including (i)~hardware-driven diversity, e.g., of antenna array size, orientation, and configuration, (ii)~device-level constraints such as varied memory and computational resources, and (iii)~environmental diversity ranging from multi-path rich urban canyons to rural line-of-sight (LOS) scenarios, and others. 
Taken together, these heterogeneity dimensions create a combinatorial explosion of configurations.

A fundamental issue is to determine the fine line between performance losses caused by inefficient application of ML models (e.g., in feature selection or unrepresentative training data) and failures that reflect their inherent limitations in handling heterogeneity.
Models often assume a fixed input-output structure tailored to a fixed BS configuration, user equipment (UE) type, and propagation environment. As a result, they fail to generalize well when applied to new environments. \textit{Domain shift}, where the input distribution between training and testing changes (e.g., due to rural/urban environments), represents only one dimension of a broader generalization problem. Practical systems often encounter \textit{concept drifts}, where the underlying input-output relationship changes (e.g., due to different codebooks) and \textit{out-of-distribution} (OOD) inputs (e.g., with previously unseen hardware configurations). 
As a result, many ML-based beam management approaches perform well in controlled homogeneous simulation settings but require separate models for each realistic configuration, which may become impractical to scale.

While some academic work acknowledges domain shift, it still remains unclear whether unsuccessful generalization in wireless systems stems primarily from poor methodological choices or happens due to deeper theoretical barriers. 
In this article, we argue for recognizing heterogeneity as a central and under-addressed barrier to the practical deployment of ML-based beam management. We analyze the key dimensions of heterogeneity in highly directional beam management, including variation in antenna configuration, codebooks, propagation environment, and user device capabilities. 
\rev{Our contributions are threefold: we (i) introduce a taxonomy of heterogeneity dimensions and discuss their effect on ML-assisted beam management, (ii) present empirical case studies that demonstrate their impact on the system-level generalization, and (iii) suggest research directions that enable heterogeneity-aware design in practical beam management systems.}

\section{Data-Driven Beam Management in 5G/B5G}
Beam management is the task of establishing and maintaining directional links between the UE and base station (BS) through beam sweeping, measurement, selection, and reporting. 

\subsection{Beam Management Procedures in 5G NR}
In 5G NR, beamforming configuration selection procedures are implementation-specific and not explicitly defined by the standard beyond timing and signaling constraints. Practical implementations typically consist of three stages.  

{First, the BS performs coarse beam sweeping.} 
The BS transmits a set of reference signals using codebooks that typically cover the entire angular space. This procedure is integrated with the initial access via transmissions of synchronization signal blocks (SSBs). \rev{SSB signals are broadcasted periodically and can be subsequently measured for maintaining synchronization, beam management, and mobility.}

{Further, the BS initiates CSI-RS-based beam refinement.} 
The BS refines its transmit beam using channel state information reference signals (CSI-RSs) and a narrower-beam codebook focused around the previously selected SSB beam. In contrast to SSB, CSI-RS transmissions can be scheduled periodically or aperiodically. 

{Finally, the UE may perform additional beam refinement.}
The UE refines its transmit beam by transmitting sounding reference signals (SRS) scheduled by the BS, periodically or aperiodically. The BS uses a preconfigured receive beam or sweeps over several beams.

Beamforming configuration selection is not limited to initial access and can be re-triggered if the UE connectivity quality (for example, estimated block error rate) is lower than the configured threshold \cite{heng2021six}, e.g., in the case of handover or beam recovery \cite{heng2024site}. 
\rev{For large beam search spaces typical for large-scale antenna arrays, beam management can induce high beam management overheads, which can be especially critical for dynamic scenarios with short beam coherence intervals.}
This motivates the use of data-driven methods that may help reduce the signaling overhead by predicting the best configuration.

\begin{figure*}
\centering
\vspace{-2em}
    \includegraphics[width=0.7\textwidth]{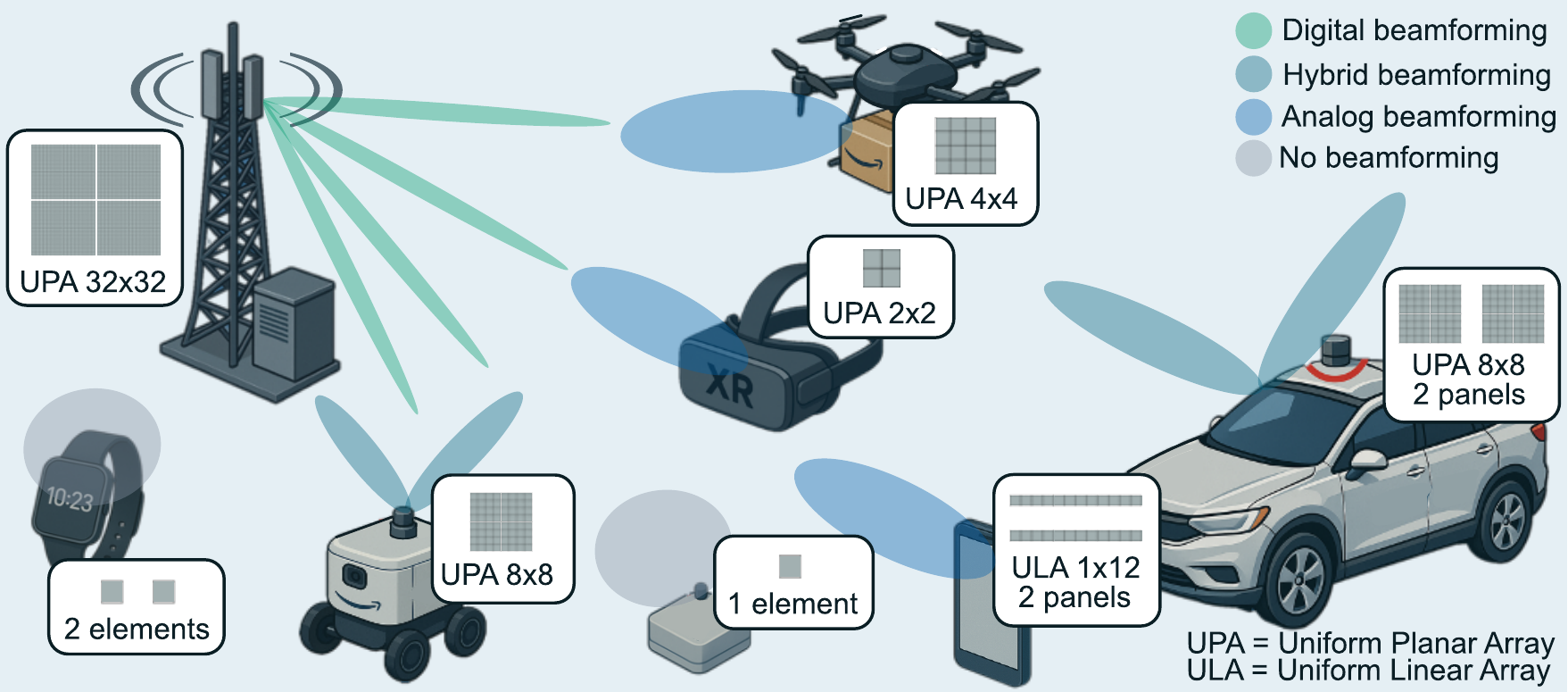}
    \caption{Illustration of heterogeneity types in data-driven beam management.}
    \vspace{-1em}
\end{figure*}

\vspace{-0.5em}
\subsection{ML Approaches for Beamforming Configuration} 
The data-driven methods commonly rely on machine learning (ML) algorithms to infer the best beamformer (constructed or selected from a predefined codebook) based on the history of previous measurements or simulations. Supervised learning and reinforcement learning (RL) are the most widely applied types of ML algorithms for data-driven beamforming. Both of them allow optimizing the beamforming configuration selection procedure based on
indicators, such as reported CSI, measured RSRP, or previously selected beams. 

Supervised learning relies on labeled data that can be collected either from real networks directly or from simulation environments. While traditionally, supervised algorithms assume offline training, the methods can also be adapted to online settings using incremental learning algorithms. Such models can deliver high predictive performance, but they often require fine-tuning to adapt to new data.  

In contrast, RL methods optimize decision policies via interaction with the environment. The environment provides a direct feedback in the form of a reward, such as throughput or signal quality.
RL-based algorithms are often pretrained offline and then deployed in the network as a fixed policy \rev{to avoid radio resource-intensive onsite policy optimization.}
Similar to supervised solutions, these methods can be pre-trained in a virtual environment and further refined on on-site measurements, if QoS requirements allow online exploration.

\subsection{Enhancing Model Generalization}
To improve model generalization, the community relies on two key strategies: fine-tuning on new data and training on diverse datasets~\cite{3gpp_tr_38_843, dreifuerst2024machine}. Fine-tuning relies on retraining the model on a small portion of device- or environment-specific data that was underrepresented or absent in the training dataset.
In practice, fine-tuning can last several minutes \cite{dreifuerst2024machine} or much longer, depending on the volume of test data and available computing resources~\cite{xu2025scenario}. During this period, the device may need to fall back on the conventional beam selection procedures. An alternative approach is to pre-train models on a mixture of data drawn from diverse measurement or simulation scenarios, which can reduce fine-tuning overheads.

Both strategies have practical limitations. On-device fine-tuning can be infeasible for resource-constrained devices due to limited compute and battery capabilities. Even if possible, slow retraining may result in 
a model outdated w.r.t. new conditions. Mixed data training may suffer from a limited number of configurations (e.g., antenna geometries, codebook sizes, environments) available for training. 

\rev{To avoid on-device fine-tuning, environment- and device-specific models can be pulled from the site-specific model catalog (e.g., during handover). Such a register, however may only contain models for the most popular devices. 
Additionally, in practice, models trained for specific devices or environments may underperform relative to matched conditions when hardware, software, or deployment conditions deviate from training assumptions. Consequently, generalization cannot be ensured by model selection alone and requires performance monitoring and fallback mechanisms. 
}
These limitations highlight that relying only on model engineering is not sufficient to ensure robustness across diverse scenarios.  

\subsection{Input Feature Space and Contextual Information}
The choice of input features plays a key role in the resulting performance of ML-based methods. Most models typically rely on physical layer measurements such as RSRP values collected from SSB and/or CSI-RS corresponding to a subset of beams or a full codebook. The measurements are often passed into a neural network-based model trained for classification (e.g., beam index prediction), regression (e.g., RSRP prediction), or RL. To enhance the performance, physical-layer measurements can be complemented with the side information, including GNSS-based locations, IMU sensor-based orientation, and information from LIDARs, cameras, or other sources. 

Simpler types of side information, such as location or orientation, are often treated as-is and either passed as additional features into the model or combined with beam measurements \cite{ali2021orientation}. High-dimensional inputs, such as images or LIDAR clouds, are preprocessed with deep encoders to generate more compact and representative features. These features are subsequently concatenated with the main inputs (e.g., beam RSRPs) or, alternatively, can be used as a context.  

While deep learning architectures successfully infer complex input-output dependencies for image, audio, or text data, their effectiveness in beamforming configuration selection tasks can be limited in the case of low-dimensional input. Moreover, these architectures implicitly encode hardware configurations and environment and therefore struggle to generalize~\cite{jayaweera20245g, 3gpp_r1_2307240}. 
\rev{This limitation is not only an implementation choice, but it reflects a fundamental coupling between commonly used features and deployment-specific parameters that cannot be fully removed by increasing model complexity.}

\subsection{Target Performance Metrics}

Model performance strongly depends on the choice of model architecture and richness of the input data. Even for well-designed models, however, selecting suitable evaluation criteria remains a key issue. Poorly chosen performance metrics may lead to inadequate choice of models. For beamforming configuration selection, traditional computer vision metrics (e.g., Top-1 accuracy) can be misleading as they ignore the resulting link-quality variation. Evaluation should instead use wireless-specific metrics such as spectral efficiency, throughput, or RSRP gap. 

A commonly used metric is the accuracy score, which can be interpreted as the frequency of selecting the strongest beam or pair of beams. Accuracy is suitable only for classification tasks, where there is only one correct label, and misclassification events should be minimized. Applied to beam prediction, the accuracy score only makes sense in environments with \textit{only one} dominant transmission direction, which often corresponds to LOS. In this case, however, \rev{it may be possible to reduce the beam search to several candidate beams or even compute the beamformers straightforwardly without any ML if the coordinates of BS and UE are known.}
Scatter-rich environments, meanwhile, can have multiple arrival directions with similar path gains. Therefore, selecting a slightly weaker transmission direction may not affect the system-level performance, making accuracy less informative.

To avoid misinterpretations, one should prioritize wireless-specific performance indicators, such as throughput, spectral efficiency, or the difference between RSRP of predicted and strongest beam directions, over traditional computer vision metrics. \rev{Deployed models should also be continuously assessed for in-field performance drift caused by changes in traffic load, propagation conditions, or hardware configuration. This can be done using a combination of system-level KPIs (e.g., throughput, BLER) and intermediate indicators (e.g., RSRP gap between predicted and measured beams or prediction confidence).} 

Wireless-specific metrics can depend on the simulation setup, such as the sizes of antenna arrays, the number of antenna panels, propagation, mobility patterns, or user density and interference. This could be resolved by standardizing simulation settings and introducing benchmark evaluation scenarios, which is an open challenge in the field.

\section{Toward Heterogeneity-Aware Beam Prediction}

\rev{Strong model generalization capabilities are essential for improving robustness of beam predictors to different heterogeneity types.} 
In this section, we outline practical considerations for enhancing the resilience of beam prediction methods to heterogeneity arising from antenna geometry, codebook design, deployment environment, and computational constraints.

\subsection{Enhancing Beam Prediction with Domain Knowledge}
The key strategy for improving the robustness of ML-based methods to hardware and environment heterogeneity is the use of wireless-specific feature representations. Prior works on generalization in beam prediction demonstrated that conventional input features such as beam RSRP are implicitly coupled with different antenna geometries, codebooks, or propagation environments and therefore lack adaptability to new configurations and deployments \cite{zhu2024physics}. This highlights the need for selecting features that remain invariant across hardware variations.

One can employ domain-specific feature representations that are minimally coupled with device- or site-specific parameters and therefore exhibit better consistency across diverse setups. The examples of such representations are beamspace representation~\cite{dreifuerst2024machine}, power-angular spectrum \cite{kilinc2024position}, and angular-delay profile \cite{vuckovic2024paramount}, which project the channel features into the angular domain. These representations are agnostic to the antenna size and shape and, therefore, can transfer effectively between different configurations.

Another promising approach is to use physics-informed models \cite{zhu2024physics} that can explicitly learn underlying channel parameters, e.g., path attenuation, angle-of-arrival, and delay coefficients from shooting-and-bouncing ray tracing simulations. Such models offer good generalization but remain computationally demanding and require accurate physical models of each deployment.
\rev{These approaches highlight the importance of aligning learned representations with the underlying physical structure of the wireless channel.}

\subsection{Antenna Geometry Mismatch}
Cellular networks typically host a diverse set of devices, including smartphones, vehicles, drones, and other platforms with distinct hardware configurations. This results in substantial heterogeneity in antenna configurations, including differences in array geometry (e.g., UPA, ULA), number of elements and active ports, panel count, and deployment orientation. While beam prediction models trained for particular antenna configurations can perform well under matched conditions, even small changes in test configurations may cause performance degradation. New antenna geometries or configurations can therefore lead to significant generalization difficulties, as demonstrated in previous works \cite{mashaal2024protobeam, jayaweera20245g}. 

For example, if a beam predictor relies solely on partial RSRP reports for a fixed set of beams, its performance is likely to degrade when applied to previously unseen antenna configurations or associated codebooks.

\subsection{Impact of Codebook Design}
The 5G NR codebook design is not specified by the standard, except for CSI reporting. As a result, BSs and UEs can employ a wide range of codebooks, ranging from 3GPP-specified CSI reporting codebooks to site-specific or ML-generated ones. These codebooks may differ in both structure and size (e.g., oversampling factor), beamwidth, and angular coverage, which makes it critical to design beam prediction solutions that can generalize to previously unseen codebooks.

\rev{From the perspective of ML-based beam prediction, changes in codebook design directly affect discretization of the angular domain and mapping of physical directions into beam indices. Predictors that operate on beam indices or other codebook-dependent measurements implicitly assume a fixed structure. 
When the codebook changes, this correspondence and learned correlations may no longer hold, even if the underlying antenna hardware and propagation environment remain unchanged. This effect is particularly pronounced for predictors trained on sparse codebooks, where neighboring beam indices may not represent similar spatial regions after the codebook changes. Consequently, robustness to codebook diversity requires feature representations that are \emph{decoupled} from specific beam indexing and instead reflect the underlying channel.}

\subsection{Environmental Diversity}
Beam prediction models are commonly trained on measurements that implicitly depend on site-specific parameters, such as the location of reflectors, blockages, and surrounding geometry. As a result, trained models may overfit to environment-specific patterns, especially when aided by side information such as location or RSRP measurements. Consequently, when a model is deployed in a previously unseen environment, the UE may experience significant performance degradation due to a mismatch between learned dependencies and the new propagation conditions.

To mitigate this issue, one option is to locally fine-tune the model using site-specific measurements collected by the UE. This requires sufficient computational and energy resources, which may only be available for highly capable devices, such as autonomous vehicles. Alternatively, the network may store multiple pre-trained models for different environments and transfer a suitable model version to the UE as it enters a new site. Such environment-specific personalization introduces additional challenges, including transmission overheads, depending on model size, and increased storage and maintenance requirements at the network side, particularly when UE beam predictors are not device-agnostic. These challenges highlight the importance of explicitly accounting for environmental heterogeneity when designing robust ML-based beam prediction solutions.

\subsection{Differences in Computational Capabilities}
\rev{Beyond representational robustness, practical deployment of ML-based beam prediction is further constrained by heterogeneous computational capabilities across UE devices.}
UE devices differ in form factor and range from constrained smartphones to high-end platforms like autonomous vehicles with on-board GPUs. While advanced devices can run deep neural models, many UEs cannot support their compute and memory demands. 
If left unaddressed, this issue may lead to uneven access to ML-based solutions, especially for resource-limited UEs. In addition, running deep learning-based models can induce increased compute and, therefore, energy use compared to conventional beam management procedures. As these procedures are substituted with their ML counterparts, repeated inference from a resource-intensive model can lead to rapid battery depletion. 

Another critical concern is the inference latency. The inference time should be consistent with hard timings of beam management protocols. As pointed out in \cite{dreifuerst2024machine}, the channel estimation and CSI feedback procedures typically take $5$\,-\,$20$\,ms. It is, however, generally difficult to develop a neural network-based architecture for site-specific codebook design with $5$\,ms inference time and $24$\,GB VRAM \cite{dreifuerst2024machine}. While tolerable for BS, these constraints can be infeasible for UEs, most of which are often smartphones with compute- and memory-limited SoCs (system-on-chips). In addition, high computational complexity limits the capabilities for on-device retraining. 
The ability to perform device personalization varies, and some devices may lack even basic support for model updates. The efficiency of retraining is especially important, as UE may end up staying in a fallback, non-ML regime for data collection. In dynamic scenarios, prolonged retraining can result in an outdated model.

\begin{figure}
    \centering
        \includegraphics[width=0.45\textwidth]{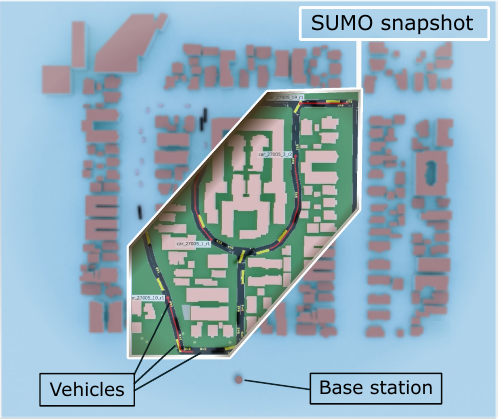} 
    \caption{3D layout of simulation scenario and SUMO snapshot.}
    \vspace{-1em}
    \label{fig:map}
\end{figure}

\section{Case Study: ML-Aided Beam Alignment}

\begin{figure*}
    \centering
    \subfloat[Antenna array heterogeneity.]{
        \includegraphics[width=0.33\textwidth]{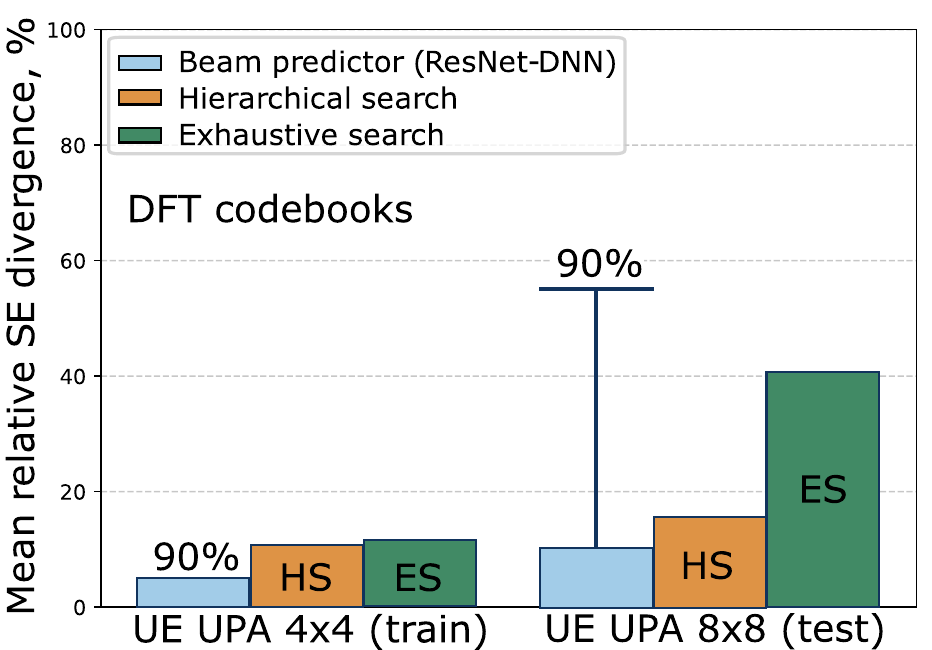}
    }
    \subfloat[Codebook heterogeneity.]{
        \includegraphics[width=0.33\textwidth]{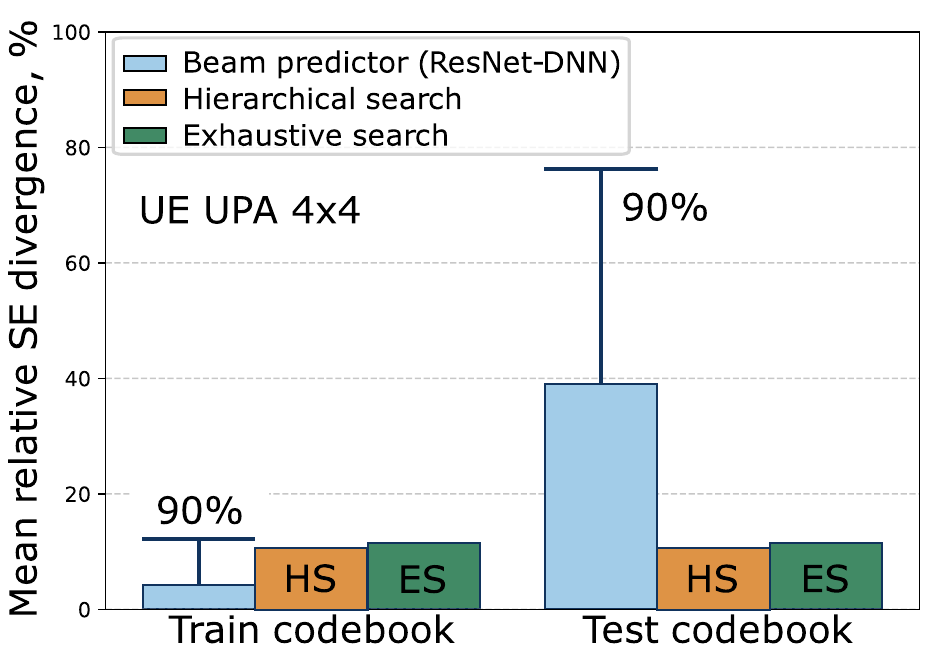}
    }
    \subfloat[Location heterogeneity.]{
        \includegraphics[width=0.33\textwidth]{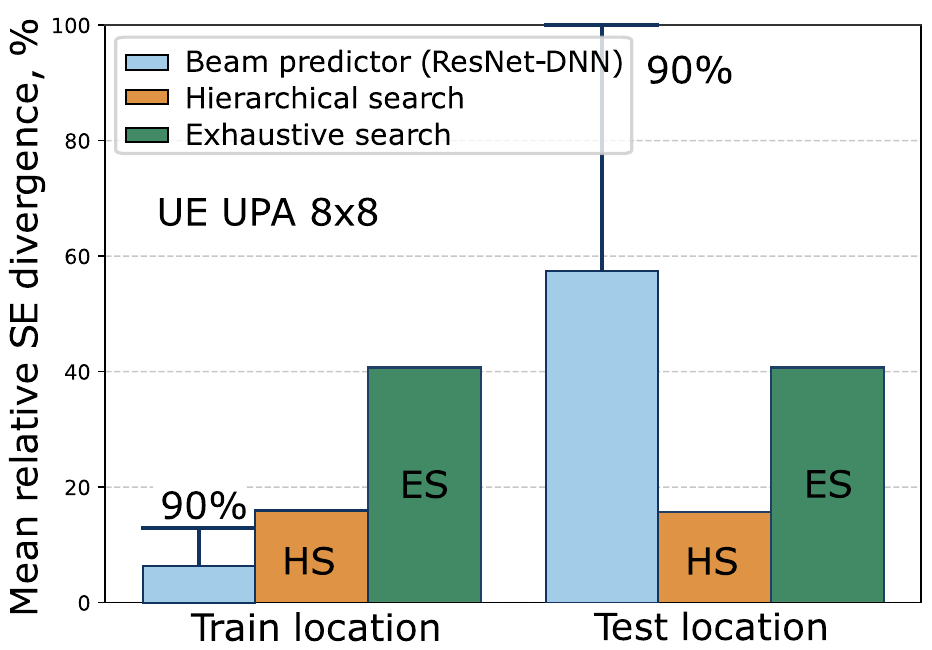}
    }
  
    \caption{Demonstration of performance drop under different UE heterogeneity types. The bars indicate the sample \emph{mean} of the SE drop \textit{relative} to the genie-based SE (thus, in \%), and the 90th percentile illustrates the variation in the distribution of the relative SE values. In the ``train'' setup, the DNN-based beam predictor (blue) always performs on par or better than two-level HS (orange) and ES (green). Under heterogeneous ``test'' conditions, however, the DNN-based predictor exhibits pronounced performance degradation and high variability, particularly under codebook and location heterogeneity. }
    \vspace{-1em}
    \label{fig:allfigures}
\end{figure*}

This section presents a numerical illustration of the effects of antenna, codebook, and environment heterogeneity on ML-aided beam selection. 

\subsection{Modeling Setup and Algorithms}

The case study uses an urban vehicular scenario with multiple mobile UEs equipped with uniform planar arrays (UPAs) in both LOS and NLOS conditions (see Fig.~\ref{fig:map}). The BS is mounted at a height of 15\,m and equipped with an $8\times8$ UPA and performs analog beamforming in a narrowband single-user MIMO setup. The UEs are heterogeneous and can be equipped with UPAs of different sizes, employ different codebooks, or reside in different locations.
All UE antennas are configured such that the broad side faces upward, which models the deployment on a vehicle where the array is parallel to the ceiling. We select a carrier frequency of 15\,GHz, 24 subcarriers, and 30\,kHz subcarrier spacing. Beam selection at both ends is performed using analog beamforming codebooks. \rev{In the considered scenario, the BS always predicts the best beam, thus focusing on the UE-side beam prediction and isolating generalization effects at the receiver}. The transmit power is 20\,dBm, and the receiver noise figure is 10\,dB. To reflect realistic city dynamics, mobility is modeled in the Eclipse SUMO simulator with a mixed urban traffic pattern including both cars and buses (3:7 ratio) at the average speed of 9.5\,m/s).  

To evaluate beam prediction performance, we adapt a ResNet-based deep neural network (DNN) architecture from~\cite{3gpp_r1_2307240, jayaweera20245g}. 
The DNN predicts received signal power along a beam direction (BD) represented by the mean azimuth and elevation. 
As non-ML baselines, we select (i) two-level hierarchical search (HS) and (ii) exhaustive search (ES), where overheads affect the resulting performance. 
In all experiments, we measure the \emph{relative} average spectral efficiency (SE) drop w.r.t. the performance of a \emph{genie-based} algorithm (i.e., 0\% corresponds to the exact genie performance). This metric directly reflects the system-level impact of beam misprediction discussed in Section III. 
To provide more insight into the distribution of relative SE drop, we also estimate the 90\%-quantile to show where 90\% of measurements fall. 

\subsection{Diversity of Antenna Configurations}

We illustrate the impact of antenna configuration diversity on the relative SE divergence w.r.t. the genie-based beam selection in Fig.~\ref{fig:allfigures}(a). In this experiment, the UEs train the DNN-based beam predictor on a $4\times4$ (``train'') UPA configuration and evaluate it on an $8\times8$ (``test'') configuration. \rev{For both configurations, UEs adopt DFT codebooks with antenna-dependent sizes of $16$ and $64$ beams for $4\times4$ and $8\times8$ UPA, respectively.} 
The color code (blue/orange/green) corresponds to the beam selection methods (DNN-based/HS/ES). For each method, we provide a 90th percentile of the SE drop to demonstrate the worst-case performance in 90\% of cases. Lower percentiles indicate more consistent performance. Non-ML baselines demonstrate stable behavior, and the distribution is concentrated closely around the mean value. Therefore, the measured empirical percentiles coincide with the mean in the plots. 

The DNN-based algorithm shows comparable mean SE divergence when transitioning from $4\times4$ to $8\times8$ UPA. The variance, however, increases significantly. 
The high 90th percentile suggests that 90\% of the measurements can suffer more than 50\% SE reduction compared to the genie-based solution. 
This behavior suggests that ML-based beam predictors trained on a fixed antenna configuration may offer limited robustness when deployed on larger arrays and become less preferable than the conventional non-ML-based beam selection algorithms.

\subsection{Codebook Type Heterogeneity}

To study this effect, we select UEs equipped with $4\times4$ UPAs and construct two distinct codebooks by randomly sampling 16 beamformers from a $\times4$ oversampled DFT codebook. 
The model is trained on one codebook and evaluated on the other. The results in Fig.~\ref{fig:allfigures}(b) show the generalization performance of the BD-based method when transferring to a new codebook. The high divergence of the DNN-based predictor indicates that this method does not achieve robust, codebook-agnostic prediction, as reflected in both the average and 90-percentile performance. Although BD-based regression in general is inherently flexible to new codebooks, these results suggest that using beam direction alone may be insufficient to bridge the domain gap introduced by changes in the codebook design. 

\subsection{Diversity of Environments}

We illustrate the effect of environmental diversity through an experiment in which the simulation environment is divided into four spatial quadrants. In this set of experiments, we compare beam predictors trained on the measurements for $8\times 8$ UPA. The train location includes one upper-right quadrant, while the test location includes the other three quadrants. The results shown in Fig.~\ref{fig:allfigures}(c) indicate a notable drop in SE when the test quadrant is not seen during training. While the performance difference between the ML-based and non-ML-based (ES/HS) changes less dramatically than in the previous scenario, a noticeable degradation of the DNN-based approach is still observed in the ``test'' setup. 

\rev{The presented results demonstrate that ML-aided beam selection perform well under matched training and deployment conditions, but its generalization degrades in the presence of antenna, codebook, and environmental heterogeneity. These effects are also reflected in increased variability and worst-case behavior, which are critical factors for reliable system-level design. These observations indicate that under device heterogeneity, robustness cannot be achieved solely through extensive model tuning or increased training data. Instead, there is a clear need for the development of learning approaches that explicitly account for physical structure, deployment diversity, and practical constraints. Building on these observations, we outline several promising future directions below.
}

\section{\rev{Future Directions and Discussion}}
The identified failure modes motivate several research directions beyond increasing model complexity and data corpus.
\subsection{Lightweight Learning for Regularization}
One of the solutions to address both distributional mismatch and computational heterogeneity discussed in Section III-E is to rely on simpler models that may act as implicit regularizers in heterogeneous deployments. For example, SVM offers analytically rigorous generalization guarantees and can robustly learn from a very limited number of samples~\cite{hua2023lab}. Combined with carefully crafted system-aware input, simple models such as SVMs or even decision trees can outperform overparameterized deep architectures relying on raw inputs. Such models are particularly attractive for resource-constrained devices, where training or fine-tuning deep models is infeasible. 
Incremental implementations of SVM also support fast adaptation in dynamic environments.

Computational heterogeneity can also be addressed by decreasing inference complexity with well-established techniques, such as model quantization, distillation, or pruning. Although these techniques do not resolve distributional mismatch, they primarily help reduce deployment barriers. Quantization compresses model weights into lower-precision formats, which allows reducing computation at the cost of potential numerical instability or reduced retraining flexibility for on-device retraining. 
Distillation transfers knowledge from a large ``teacher'' model to a smaller ``student'' model, which helps improve inference latency and memory usage. Yet, distilled models can lose their performance when fine-tuned due to their smaller capacity, which may render them fragile in dynamic or highly diverse deployments. Further, pruning removes model weights with marginal contribution to overall performance, but if aggressive, it may degrade accuracy or require additional retraining. 

\vspace{-0.5em}
\subsection{Collaborative Learning for Knowledge Exchange}

Distributional mismatch across devices motivates the need to exchange information about beam-channel relationships and, therefore, the use of collaborative learning techniques. While federated learning (FL) enables collaborative model update without sharing raw data, it does not inherently resolve the domain shift caused by diversity, and in the presence of heterogeneity, can even lead to performance degradation if local data distributions are highly non-IID.
Model-level personalization allows each device to maintain a shared global model and a personalized component. The shared component captures features universal for the majority of devices or environments, while the personalized part adapts to device- or environment-specific properties.

Furthermore, diversity of sensors results in different types of side information, which further complicate collaboration. 
Devices equipped with sensors with rich contextual inputs (e.g., LIDAR) generate updates that are not directly transferable to devices that rely only on low-dimensional inputs such as location or orientation. 
This can be resolved by combining devices into groups according to the device-specific data and hardware profiles %
and training models within such groups, which may become costly.
Another challenge is to mitigate the effect of so-called stragglers (i.e., devices with weak compute powers violating timing constraints). Straggler-aware strategies, such as asynchronous updates, sampling, or resilient aggregation techniques, can improve timing alignment of compute-limited devices. Overall, collaborative learning, %
carefully crafted beyond simple gradient averaging, can reduce the need for local adaptation.

\vspace{-0.5em}
\subsection{Adaptive Learning}
Resilience to dynamic conditions and deployment drift requires the design of adaptation mechanisms compatible with the operational constraints of beam management (e.g., latency, energy, etc.). Adaptation should be lightweight and driven by system performance indicators.
One such form of adaptation for a constant beam-channel structure is continual learning. Continual learning enables models to incrementally adapt to new conditions, such as changes in mobility patterns or environments, without retraining from scratch~\cite{mohsin2025continual}. 
At the same time, while ML models can be adapted to environment dynamics, in the presence of structural changes (e.g., codebook mismatch), na\"ive continual updates may fail to adjust to new conditions.

Active learning, where devices can be specifically requested to provide labeled data when confidence degrades, provides an additional adaptation mechanism. 
This may correspond to requesting expanded beam sweeps or higher-resolution measurements in specific conditions. 
While such strategies can improve efficiency, they need to be carefully designed to avoid excessive signaling overhead or operation disruptions.

Summarizing, these observations indicate that while part of the generalization challenge arises from how ML is applied, others may reflect deeper structural limitations imposed by hardware or sensing diversity. Such limitations cannot be reliably resolved through more complex models and data alone. Recognizing these and designing accordingly represents the key research direction for ML-aided beam management.
Future work should focus on developing theoretical tools to characterize the ability for generalization under heterogeneity and on establishing standardized benchmarks for corresponding evaluations.

\bibliographystyle{ieeetr}
\bibliography{references.bib}

\begin{IEEEbiographynophoto}{Nikita Zeulin} is a Doctoral Researcher currently pursuing a doctoral degree with Tampere University, Finland. He received his B.Sc. degree from St. Petersburg State University of Aerospace Instrumentation, Russia, and his M.Sc. degree from Skolkovo Institute of Science and Technology, Russia. His research interests 
include federated learning and machine learning enhancements for wireless communications.
\end{IEEEbiographynophoto}

\begin{IEEEbiographynophoto}{Olga Galinina} 
is an Institute of Advanced Study Senior Research Fellow at Tampere University, Finland. She received her B.Sc. and M.Sc. degrees from the Department of Applied Mathematics, St. Petersburg State Polytechnical University of Peter the First, and her Ph.D. degree from Tampere University of Technology. Her research interests include applied mathematical modeling and analysis of wireless networks and statistical machine learning.
\end{IEEEbiographynophoto}

\begin{IEEEbiographynophoto}{Ibrahim Kilinc}
is a Ph.D. student at the University of California, San Diego. He received his B.S. in Electrical and Electronics Engineering and his minor degree in Information Systems and Technologies from Bilkent University, Ankara, Turkey, in 2023. His research interests include sensor-aided beam alignment in heterogeneous devices in wireless systems and signal processing.
\end{IEEEbiographynophoto}

\begin{IEEEbiographynophoto}{Sergey Andreev} received the Ph.D. degree from Tampere University of Technology, and the Cand.Sc., and Dr.Habil. degrees from SUAI. He is a Professor of Wireless Communications and formerly an Academy Research Fellow with Tampere University, Finland. He is also a Research Specialist with Brno University of Technology, Czech Republic.
He co-authored more than 300 published research works on intelligent IoT, mobile communications, and heterogeneous networking.
\end{IEEEbiographynophoto}

\begin{IEEEbiographynophoto}{Robert W. Heath Jr.}
is the Charles Lee Powell Chair in 
Wireless Communication in the Department of ECE at the University of California, 
San Diego. He has received several awards including the 2019 IEEE Kiyo Tomiyasu 
Award, the 2020 North Carolina State University Innovator of the Year Award, and 
the 2021 IEEE Vehicular Technology Society James Evans Avant Garde Award. He 
is a licensed Amateur Radio Operator, a registered Professional Engineer in Texas, 
a Private Pilot, a Fellow of the National Academy of Inventors, a Fellow of the IEEE, 
and a Fellow of the AAAS.
\end{IEEEbiographynophoto}

\end{document}